\documentclass[twocolumn]{revtex4-1}

\usepackage{graphicx,dsfont,color}
\usepackage[export]{adjustbox}

\begin{document}

\title{Quantum Non-Gaussian Multiphoton Light}

\author{Ivo Straka\textsuperscript{\textdagger}}
\email[Corresponding author: ]{straka@optics.upol.cz}
\author{Luk\' a\v s Lachman}
\thanks{These authors contributed equally to this work.}
\author{Josef Hlou\v sek}
\author{Martina Mikov\' a}
\author{Michal Mi\v cuda}
\author{Miroslav Je\v zek}
\author{Radim Filip}
\email[]{filip@optics.upol.cz}

\affiliation{Department of Optics, Palack\' y University, 17.\ listopadu 1192/12,  771~46 Olomouc, Czech Republic}

\begin{abstract}
We propose an experimental method of recognizing quantum non-Gaussian multiphoton states. This is a native quantum property of Fock states, the fundamental quantum states with a constant number of particles. Our method allows experimental development and characterization of higher Fock states of light, reaching even beyond the current technical limits of their generation. We experimentally demonstrate that it is capable of distinguishing realistic quantum non-Gaussian light with the mean number of photons up to 5 despite detection efficiency of 50 \%. We also provide evidence that our method can help to distinguish the number of single-photon emitters based only on their collective emission.
\end{abstract}

\maketitle

\section{Introduction}

\begin{figure*}[ht!]
\begin{tabular}{c@{\hspace{10mm}}c}
\includegraphics[width=.47\linewidth,valign=t]{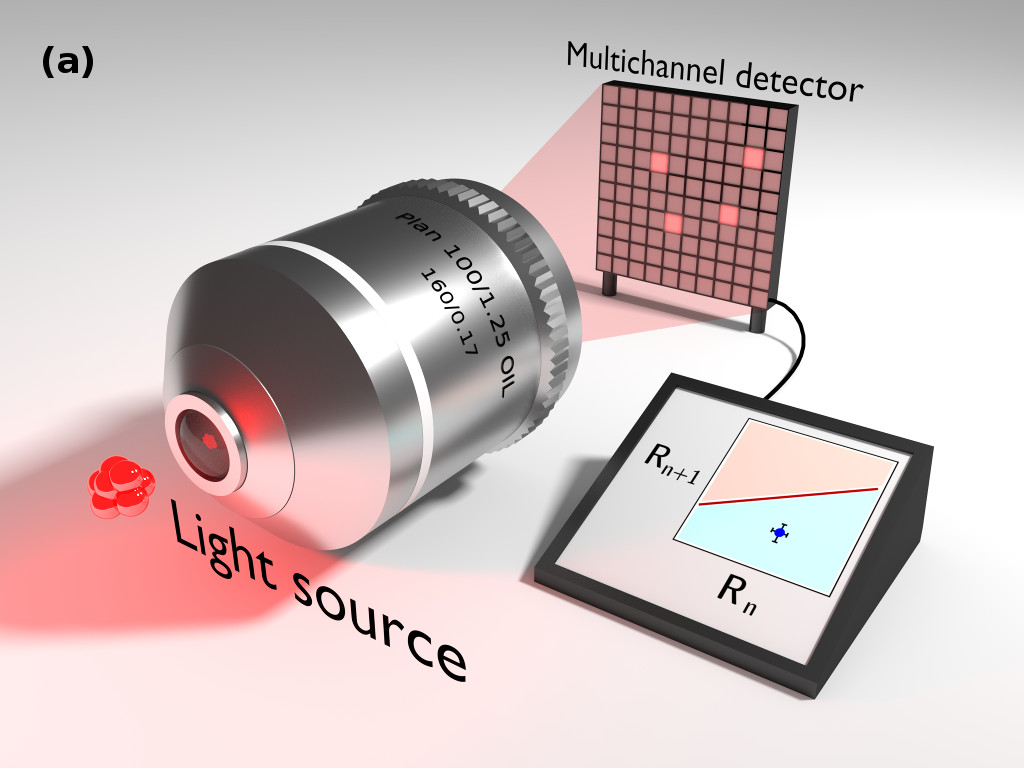}
&
\includegraphics[width=.47\linewidth,valign=t]{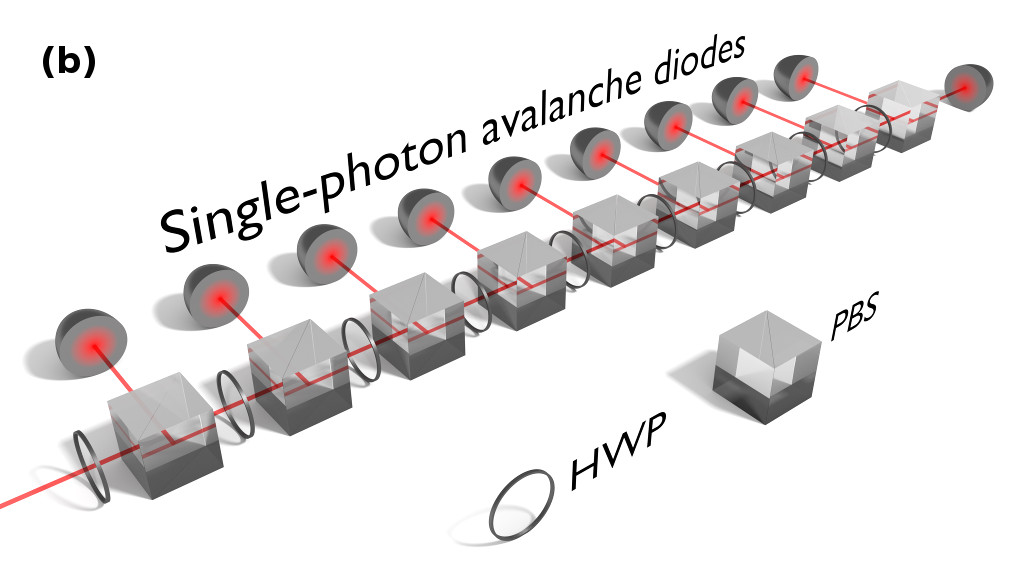}
\end{tabular}
\caption{A general proposal of the experimental QNG witness. \textbf{(a)} Multiphoton light is collected and brought to a balanced multichannel detector, where coincidences $R_n,R_{n+1}$ are compared to the QNG threshold. \textbf{(b)} The detector in our experiment consists of 10 silicon single-photon avalanche diodes (SPAD) and a balanced array of polarizing beam splitters (PBS) and half-wave plates (HWP) to control the splitting ratio. The half-wave plates can be adjusted to split the light equally to any number of selected channels, so there is no need to physically add or remove SPADs.}
\label{fig.scheme}
\end{figure*}

Over the past few years, generation of multiphoton nonclassical states of light became important for quantum technology \cite{NOONPhase,GHZ}. Their unique feature is that they represent a bundle of indistinguishable photons with pronounced particle behavior. This behavior is altogether incompatible with any semi-classical continuous optical waves \cite{WangGaussianStates,BraunsteinVanLoock,ParisGaussianStates,LloydGaussianStates}. Many recent experimental steps brought us closer to generating Fock states $|n\rangle$ of light, essential resources with a fixed and constant number of photons $n$ \cite{Lvovsky2Photons,Cooper3Photons,Furusawa3Photons}. In our discussion, we focus on the quantum aspects of Fock states. Some of them are very susceptible to optical losses \cite{FockLoss,NOONLoss}. Namely, the negative regions in their Wigner quasiprobability density are very sensitive even for ideal noiseless states \cite{Zurek}. Consequently, Fock states of light have been experimentally approached only up to $|3\rangle$ in terms of this negativity \cite{Cooper3Photons,Furusawa3Photons}. On the other hand, nonclassical light can be produced rather easily for much higher number of photons by squeezing or conditioning  \cite{Silberhorn50Photons,LeuchsBrightSqueezedVacuum}. Apparently, there is a large unexplored extension of multiphoton nonclassical states towards high Fock states of light.

Exploring this large extension of nonclassical states is a challenging task, on both theoretical and experimental side. However, quantum technology is boosting this ongoing investigation. A distinct step into the large gap between nonclassical states and Fock states was done when a criterion for single-photon quantum non-Gaussian (QNG) light was proposed \cite{QNGTheory}. Analogously to the definition of nonclassical light, QNG state cannot be expressed as \emph{any probabilistic mixture of Gaussian states}. That covers all possible sources available from linearised quantum dynamics of light. More QNG criteria utilizing phase-space quasiprobability distributions have been proposed \cite{QNGWigner,QNGsParam,QNGPark,QNGMarginal}. Experimental verification was done using both single-photon detectors and homodyne detection \cite{QNGExperiment,AnaQNGDot,QNGDepth,JezekQNGSqueezed,QNGUpconverted,Huntington}. Single photons were tested for robustness up to 20 dB of optical attenuation. Moreover, QNG has unique properties that have been explored in detail for single-photon states. Among these properties are dependence on optical loss \cite{QNGDepth} and a positive upper bound on the Wigner function at zero \cite{QNGWigner}. These two properties represent a significant distinguishment from both nonclassicality and Wigner function negativity, opening up a new class of multiphoton quantum states.

Recently, quantum non-Gaussianity of light has been proposed as a test bed for optical links suitable for quantum key distribution \cite{QNGQKD}, as opposed to the photon autocorrelation function that is typically measured. However, this single criterion \cite{QNGExperiment} cannot be efficiently applied to all multiphoton states. For example, even ideal multiphoton Fock states $|2\rangle$ to $|6\rangle$, when attenuated below the respective transmittances $\eta=0.30,0.42,0.50,0.53,0.63$, cannot be detected as QNG light using the single-photon criterion \cite{LachmanQNGEmitters}. These ideal states are, however, QNG regardless of the attenuation, so such criteria should exist.

Here we consider commonly available multichannel detectors and derive the QNG criteria precisely for them, and experimentally verify QNG multimode light up to 9 heralded photons (see Results) despite collection and detection loss of 50 \%. Although these states are apparently sub-Poissonian, our approach to QNG is not related to the photon number variance, but to experimental indivisibility of $n$ photons. This fundamental particle property limits the probability of all $n+1$ channels in the detector firing at once. This can be measured directly and does not require reconstruction or tomography of any kind. We can further estimate the robustness of multiphoton states of light under optical loss unavoidable in potential applications. The resulting loss tolerance is quantified as QNG depth, defined as the maximum optical attenuation applicable on a specific quantum state, after which the QNG can still be witnessed.

The key contribution of our proposed approach is to development of experimental multiphoton states towards ideal Fock states, as QNG represents a necessary milestone along the way, interposing between nonclassicality and Wigner function negativity. It also provides a physical measure of robustness -- QNG depth -- directly as a result. Therefore it serves both as QNG witness and quantifies the maximum tolerable optical loss.

\section{Results}
\subsection{Multiphoton QNG criteria}

A suitable case to consider is a multichannel detector, as it is widely available in many laboratories \cite{AchillesMultichannel,FitchMultichannel,WalmsleyMultichannel}. The detector splits incoming light to multiple separate single-photon binary detectors, as depicted in Fig.~\ref{fig.scheme}. We consider only detection probabilities denoted as $R_n$ and $R_{n+1}$ for the detector with $n+1$ identical channels. First, we choose $n$ particular channels, then $R_n$ is the probability of successful detection on these, irrespectively to the state of the odd channel. $R_{n+1}$ is the probability of all $n+1$ channels registering photons. Thus, we will be directly witnessing whether the measured $R_n,R_{n+1}$ are incompatible with any mixture of Gaussian states of light. Such detection technique is not sensitive to phase properties of light and can also be applied to multimode structures of light. Therefore, it is suitable for testing a wide range of optical sources.

We consider a linear functional
\begin{equation}\label{func}
F(a)=R_n+aR_{n+1}
\end{equation}
where $a$ is a free parameter. Since $R_{n,n+1}$ are linear functions of quantum states and their photodistributions, for a given $a$, there is a maximum $F_\mathrm{max}(a)$ that can be reached among mixtures of Gaussian states; that is squeezed coherent states \cite{QNGTheory}. Then, for given measured values of $R_{n,n+1}$, we get the function $F_\mathrm{meas}(a)$. If there exists any $a_0$, for which the maximum of Gaussian mixtures is surpassed, $F_\mathrm{meas}(a_0) > F_\mathrm{max}(a_0)$, the measured state is quantum non-Gaussian.

Since (\ref{func}) is a linear functional of the quantum state of light, the maximum $F_\mathrm{max}$ is reached for a pure Gaussian state \cite{Lachman2013,QNGTheory}. However, each mode is still parametrized by the maximal amplitude of displacement, minimal quadrature variance and a phase shift between them. The resulting minimum threshold for $R_n$ as a function of $R_{n+1}$ is a highly nonlinear function of three parameters per mode. For single-mode light, precise numerical solution is possible. The details of derivation are given in Methods and the Appendix. For multi-mode light, the number of parameters increases, so we used Monte-Carlo simulations to find the thresholds. As is shown in the Appendix, the multi-mode thresholds are identical to the single-mode thresholds.

For light with small mean number of photons, the thresholds can be approximated by a useful analytical formula 
\begin{equation}\label{analyt}
R_{n}^{n+2}>H_n^{4}(x) \left[ \frac{R_{n+1}}{2 (n+1)^3}\right]^{n},
\end{equation}
where $H_n(x)$ is the maximum value of a Hermite polynomial among such $x: H_{n+1}(x)=0$. Derivation is presented in the Appendix where a more precise approximation is also derived.

\subsection{Experimental Results}

\begin{figure*}[ht!]
\begin{tabular}{c@{\hspace{10mm}}c}
\includegraphics[width=.38\linewidth,valign=t]{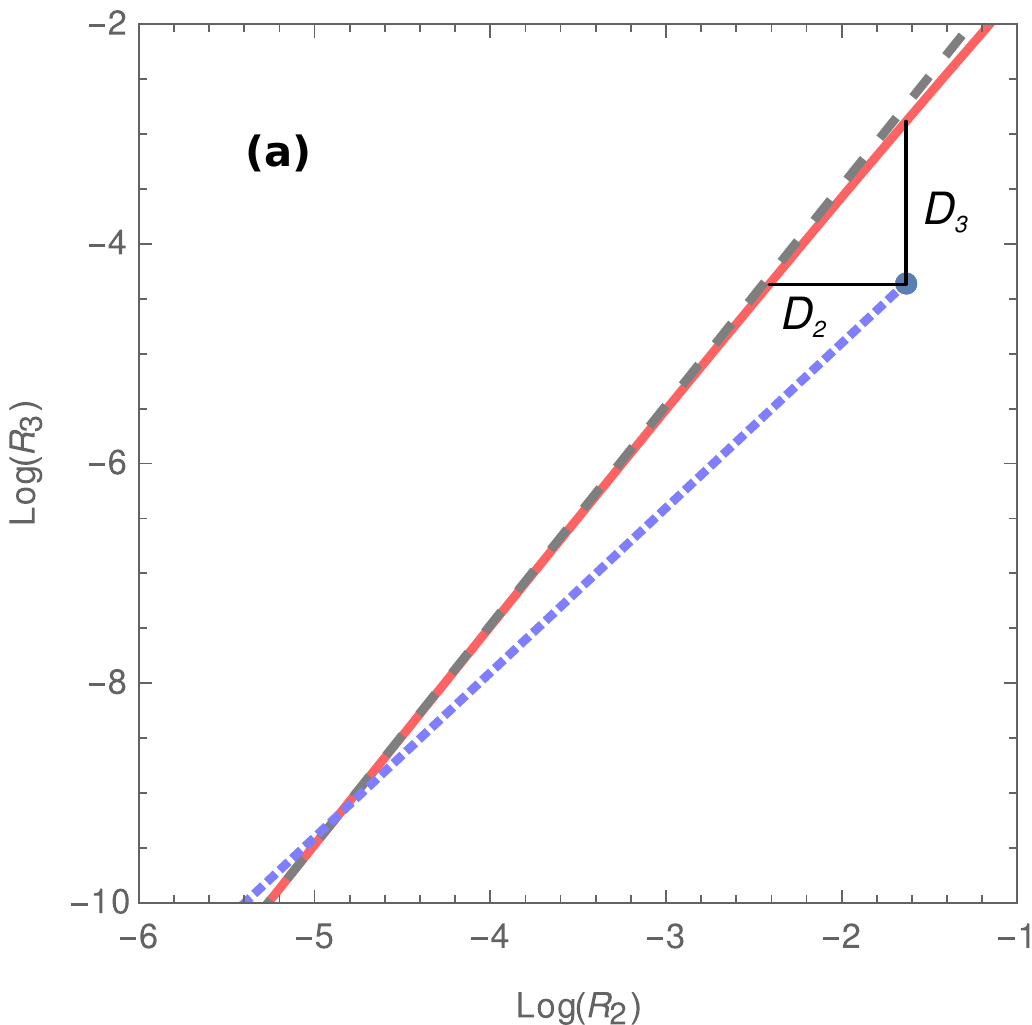}
&
\includegraphics[width=.38\linewidth,valign=t]{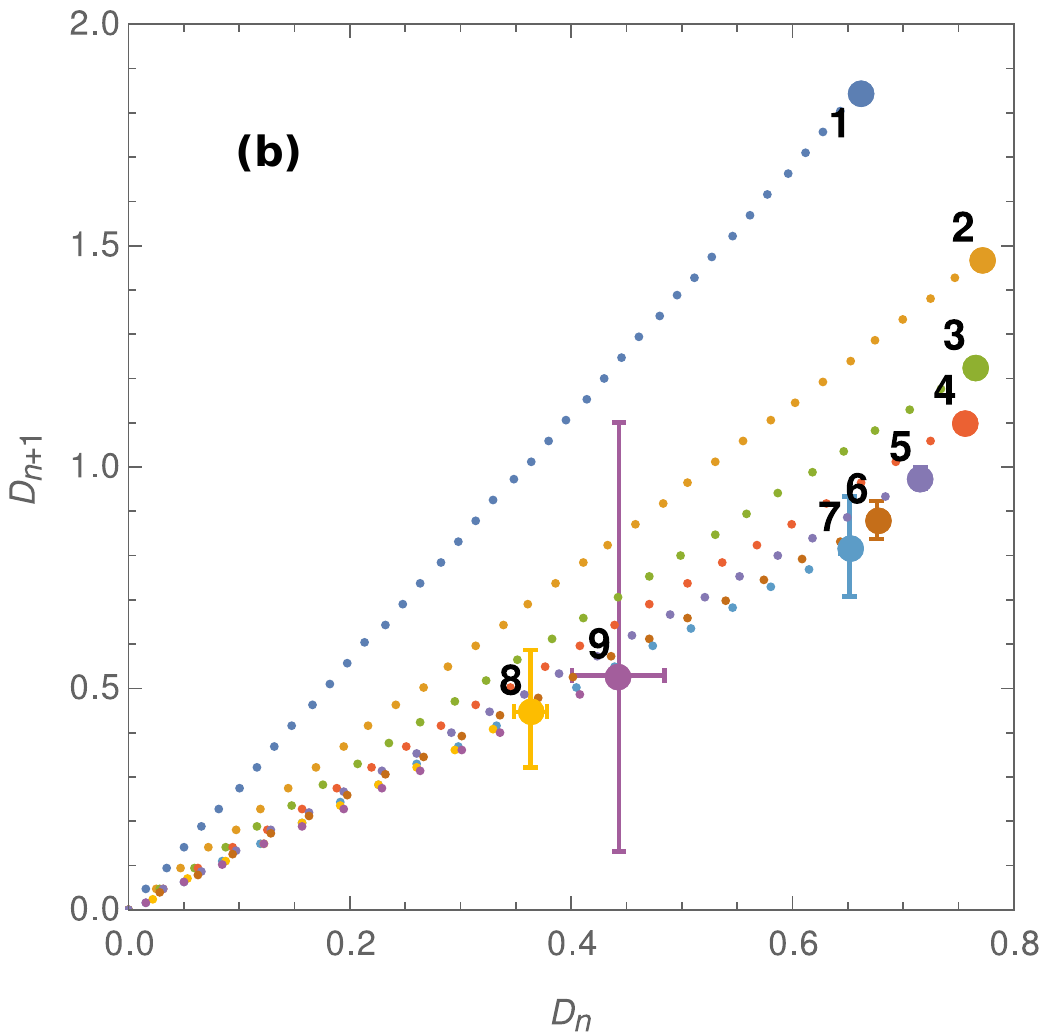}
\end{tabular}
\caption{QNG tests for heralded 1-9 photons. \textbf{(a)} Example for $n=2$. Red line represents the QNG criterion, while dashed grey line is its approximation (\ref{analyt}). The blue point represents the measured state and the dotted blue line is the path of the point if the state becomes further attenuated. For the sake of visualising all data, we denote $D_n, D_{n+1}$ as the horizontal and vertical log-distances between a measured point and the QNG threshold. \textbf{(b)} The distances $D_n$ and $D_{n+1}$ for generated $n$-photon states were measured on a balanced $(n+1)$-channel detector (numbered points for $n=1-9$). Reaching the point of origin at zero means that QNG is no longer recognizable. Dotted paths represent attenuation with steps of 0.5 dB. Note that these steps are not the same size for all $n$. The number of steps on each path is proportional to QNG depth. For QNG depth values, see Fig.\ \ref{fig.depth}. Both vertical and horizontal errorbars are shown for all data points; in some cases they are smaller than point size. For individual depiction of each multiphoton state with its respective QNG criterion, see Fig.\ \ref{fig.dataSupplement} in Appendix.}
\label{fig.data}
\end{figure*}

Experimentally, it is very challenging to generate a multi-photon quantum state that would be sufficiently close to a Fock state. Previous efforts have succeeded in generating heralded sub-Poissonian states with a high mean-photon-number \cite{BoydMultiphoton,LauratSubPoisson,ChekhovaBrightSubPoisson,Silberhorn50Photons}. However, QNG requires much more than sub-Poissonian light. For such heralded states, the main issue are systematic high-photon-number contributions coupled with optical loss in the trigger channel. The overall efficiency required to generate QNG light would have to be very close to 100~\% and thus beyond current technical capabilities. Multi-mode states allow us to overcome these limitations, and still the same QNG criteria apply -- if the measurement gives a certain detection statistic, the QNG criterion gives identical results regardless of the number of modes measured. Using this correspondence in detection statistic, we produced QNG multi-mode states to show that the proposed criteria can indeed recognize corresponding single-mode states. Such single-mode states would represent the missing link on the way towards Fock states, between nonclassicality and Wigner function negativity.

To achieve this, we produced multi-photon states by mixing $n$ single-photon states together incoherently using time multiplexing. As an additional advantage, this approach simulates incoherent mixing of signals from a cluster of $n$ identical single-photon emitters in separate modes. This is a very relevant topic, because recognizing non-classical properties of such clusters or simply counting these emitters is subject to ongoing research, which offers limited accuracy \cite{SauerCounting,SmallOrigamiCounting,NatCommOrigamiCounting,QDotCounting,IsraelQDResolution}.

In our experimental set-up, we used a collinear type-II spontaneous parametric down-conversion in a periodically poled KTP crystal, which was pumped by a narrow-band continuous-wave laser diode at 405 nm. We already showed that such sources generate very high-quality heralded single-photon states \cite{QNGDepth}. We took $n$ successive time windows, where a single photon was heralded, and joined them into a single temporal detection unit. Due to this approach and exceptional brightness and efficiency of the source, we were able to generate a photon statistics exhibiting QNG and proved it for $n$ up to 9 (Fig.\ \ref{fig.data}, for more details, see Methods/Appendix).

As a multichannel detector, we constructed a network of polarizing beam splitters (PBS) and half-wave plates to facilitate a balanced 1-to-$(n+1)$ splitter. The design is equivalent to the one depicted in Fig.\ \ref{fig.scheme}b, but we used a tree structure instead of a linear one. In each arm, a silicon single-photon avalanche diode (SPAD) was placed as a detector. Even though each SPAD has different efficiency, it is sufficient to adjust the PBS network so that the responses of all detectors are balanced. This way, the measured state is merely subjected to additional loss, but that does not create any false positives in QNG witnessing \cite{QNGExperiment}. The total number of detection ports was 10 on the measured state and one for the heralding detector.

We estimate the mean numbers of photons to be up to 5, if we take into account the detector efficiency that was $\approx 50 \% $. However, no correction has been done in the data, and our results represent direct witnessing of QNG using a lossy detector. In this regime, witnessing the negativity of the Wigner function would not be possible.

The measured states are very robust against optical loss, withstanding up to 5-20 dB of attenuation before their QNG character becomes undetectable. In previous work, we demonstrated that this QNG depth can be precisely predicted \cite{QNGDepth}.

\begin{figure}
\includegraphics[width=\linewidth]{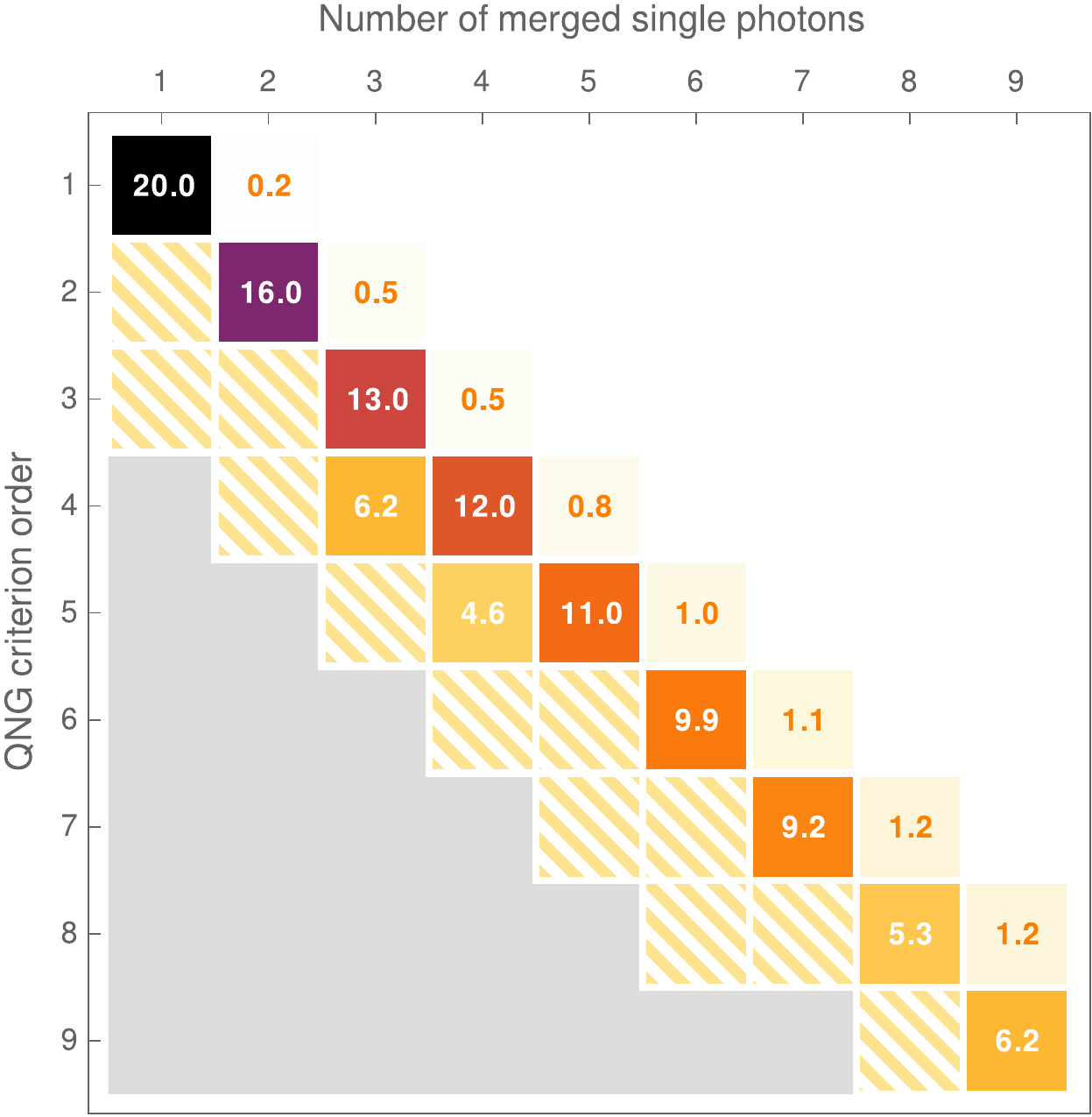}
\caption{Table of QNG depths (in dB), defined as the maximum attenuation, for which the quantum state is still QNG. The horizontal axis shows the number of single-photon states that comprise the measured state. The vertical axis represents the order of the QNG criterion used to measure the state ($n$ in Eq.\ (\ref{func})). The diagonal represents the data being shown in Fig.\ \ref{fig.data}.
Solid-colored tiles represent points with positively measured QNG despite statistical uncertainties. For points above the diagonal, the depth estimates are conservative and lower than the actual QNG depth, because $R_{n+1}$ is no longer caused solely by noise. The upper white region represents combinations of measured states and criteria that did not show QNG. Orange stripes denote measurements where statistical uncertainty intersects with the QNG criterion border, making the result inconclusive. Data in the grey region contain no detections at all.}
\label{fig.depth}
\end{figure}

In Fig.\ \ref{fig.depth}, QNG depths of various multi-photon states are shown, as measured using multiple QNG criteria. Here, each multi-photon state is positively detected with at least one order of the QNG criterion.

\section{Discussion}

Let us discuss an insight into the behaviour of the QNG criteria. As per Eq.\ (\ref{analyt}) and shown in Fig.\ \ref{fig.data}, the QNG borders are approximately linear on a log-log scale with a slope of $(n+2)/n$, the approximation being low mean number of photons. Attenuation paths behave similarly with a slope of $(n+1)/n$, assuming low realistic noise in the sense of fast-decaying photodistribution of the measured state $P_n \gg P_{n+1} \gg \sum_{k=n+2}^\infty P_k$. Because the QNG slope is always greater, the two dependencies eventually intersect and quantum states have typically some finite QNG depth. The robustness observed in our data is due to low multiphoton contributions in our single-photon states. When combining $n$ of them, the photodistribution becomes almost binomial with a weak systematic noise, $P_{\leq n} \gg P_{> n}$, and consequently, $R_n \gg R_{n+1}$ for experimental rates. This sharp contrast is necessary for QNG and was the reason for our choice of using multimode heralded single-photon states.

Generally, if the criterion order is lower than the number of merged single photons (white area in Fig.\ \ref{fig.depth}), the dominant contributions to $R_n,R_{n+1}$ arise from probabilities of heralded generation of $n,n+1$ photons, respectively. These results have a very low uncertainty, but mostly fail to pass the QNG criterion; chiefly due to optical loss \cite{LachmanQNGEmitters}. If the criterion order is higher than the number of single photons, these coincidences are always caused by noise with very low detection rates. Those cases are mostly inconclusive, unless measured for excessively longer periods of time. In between, there is always an optimal criterion that recognizes QNG for the widest range of potential optical loss (the diagonal in Fig.\ \ref{fig.depth}).

This complies well with our initial motivation to test the indivisibility of $n$ photons using a criterion with $n+1$ detectors. Practically, if we consider the measured quantum states as a simulated collective emission from $n$ identical single-photon emitters, these optima offer a potential way to count the emitters based only on their emission. This hypothesis would heavily depend on the quality of the single-photon emitters, but our data show that, at least in our simulation, QNG depth is capable of accurate distinction for a high number of emitters. Current methodology for counting or resolving individual single-photon emitters also uses multichannel detectors, but has limited accuracy. Results have been published for fluorescent dye molecules \cite{SauerCounting,SmallOrigamiCounting,NatCommOrigamiCounting} and for quantum dots \cite{QDotCounting,IsraelQDResolution}. We hope future measurements will explore this approach further using emission from physical samples.

In this work, we introduced and verified an experimental approach to direct witnessing of quantum non-Gaussian multi-photon states as a necessary step towards multi-photon Fock states. This method is capable of recognizing multi-photon states produced by multiple atoms or molecules, trapped ions or solid-state emitters, that can or cannot be individually controlled \cite{SauerCounting}. When investigating a cluster of single-photon emitters, the collectively emitted light can be detected as QNG before any single emitter is isolated or controlled. We also demonstrated that these measurements are feasible with realistic detection efficiency. Moreover, this method is generally applicable to quantum states of matter producing QNG states of light, for example in quantum optomechanics \cite{Optomechanics}.

\section{Methods}

\subsection{Experiment}

We used an emission from multiple heralded single-photon states temporally merged together. The single-photon source uses a collinear type-II spontaneous parametric down-conversion in a 6-mm thick ppKTP crystal and poling period 10 $\mu$m. As a pump, we used a narrow-band continuous-wave laser diode at 405 nm. The heralding rate was set to about 650 kHz, which corresponds to the maximum data flow allowed by the coincidence electronics. The spectral bandwidth of the photons is about 1 nm FWHM and the Hong-Ou-Mandel uncorrected visibility 98 \%. The raw heralding efficiency was about 30 \%.

The temporal merging of $n$ single-photons was achieved by considering a number of consecutive coincidence time windows as parts of one temporal detection mode. The positions of these respective time windows were heralded by a detector in one of the SPDC modes. This is equivalent to collecting light from $n$ independent single-photon emitters in $n$ modes. The criteria for $R_n,R_{n+1}$ for $n$ modes can be verified to be identical to the case of $n=1$ using the Monte Carlo approach. Additionally, the detection statistics of this state is the same as for a single-mode state with equivalent photodistribution. This only requires that all SPADs work in binary mode. To achieve this, if a SPAD registers detections in multiple coincidence windows that are part of one detection mode, it is considered as a single detection only.

For detection, we used two time-to-digital converters, each having 8 channels with the resolution of 81 ps/time bin. Since we needed 11 channels in total, we used two modules and synchronized them with a shared periodic signal at 100 kHz. This frequency was chosen to compensate the measured relative clock drift $10^5$ time bins/second.

\paragraph*{Statistical methods.}

Error bars in Fig.\ \ref{fig.data} denote minimum-width Bayesian confidence intervals (68 \%) assuming a uniform prior. The reason for this was that the primary source of statistical uncertainty was a discrete counting process, where measured values were as low as one count.

\subsection{QNG criterion}

In Eq.\ (\ref{func}), we defined a general linear functional of coincidence probabilities $R_{n,n+1}$
\begin{equation}
F(a) = R_n+aR_{n+1}.
\end{equation}
Now we show the maximization of $F(a)$ over pure Gaussian states.

The probabilities $R_{n,n+1}$ are best expressed using a probability of vacuum in an attenuated state,
\begin{equation}
P_0(\tau)=\mbox{Tr}\left[\left(|0\rangle \langle 0| \otimes \mathds{1}\right) \cdot U(\tau) \cdot \left(\rho \otimes |0 \rangle \langle 0|\right) \cdot U(\tau)^{\dagger}\right],
\label{P0}
\end{equation}
where $\rho$ is the density matrix of the state and $U(\tau)$ is a unitary operator corresponding to a beam splitter with transparency $\tau$. The probability of $n$ simultaneous detections on given channels -- $R_n$ -- is obtained by a summation
\begin{equation}
R_n=1+\sum_{k=1}^{n}(-1)^k {n \choose k} P_0(k/N),
\label{Rn}
\end{equation}
where $N$ is the total number of spatial modes, which the detector can distinguish. The identity (\ref{Rn}) implies that the probability (\ref{P0}) determines the probability $R_n$. For a pure Gaussian state (\ref{P0}) reads
\begin{equation}
P_0(\tau)=2 \frac{e^{-\frac{ \beta ^2 \tau}{2}\left[\frac{\cos ^2 \phi}{\mu(1/V)}+\frac{ \sin ^2 \phi}{\mu(V)} \right]}}{\sqrt{\mu(V)\mu(1/V)}},
   \label{P0att}
\end{equation}
where
\begin{eqnarray}
 \beta  e^{i \phi}&=&\frac{1+V}{2\sqrt{V}}\alpha +\frac{1-V}{2\sqrt{V}}\alpha^*, \nonumber \\
\mu(V)&=&2V+\tau(1-V)
\end{eqnarray}
with $\beta$ real and positive. The parameter $\alpha$ denotes a complex amplitude of a coherent state which undergoes quadrature squeezing and $V$ is the minimal quadrature variance in time.

The expressions (\ref{Rn}) and (\ref{P0att}) give raise to the linear combination of probabilities for a pure Gaussian state
\begin{equation}
F_{a,n}(\beta,V,\phi)=R_n+aR_{n+1}.
\label{rV}
\end{equation}
Thus, the optimizing task is well parametrized and can be done numerically. It has turned out that the global maximum requires $\phi=0$ for each $a$ and order $n$. The remaining two parameters fulfil a necessary condition of a local extreme
\begin{equation} \label{extremalConstraint}
\partial_{\beta} R_n \partial_{V} R_{n+1}=\partial_{V} R_n \partial_{\beta} R_{n+1}.
\end{equation}
It ensues from the exclusion of the $a$ parameter from the equations giving conditions on a local extreme. Although this relation does not specify $a$, it leads to the solution of the problem, when the task is understood equivalently as optimization of the probability of success $R_n$ over the set of states with a constraint on error probability $R_{n+1}$ and Lagrange multiplier $a$. The resulting maximal success probability is bound to the error probability only by a single parameter determined by the relation (\ref{extremalConstraint}).

\section{Appendix}

\subsection{Quantum non-Gaussianity criteria}

\begin{figure*}
\includegraphics[width=\linewidth]{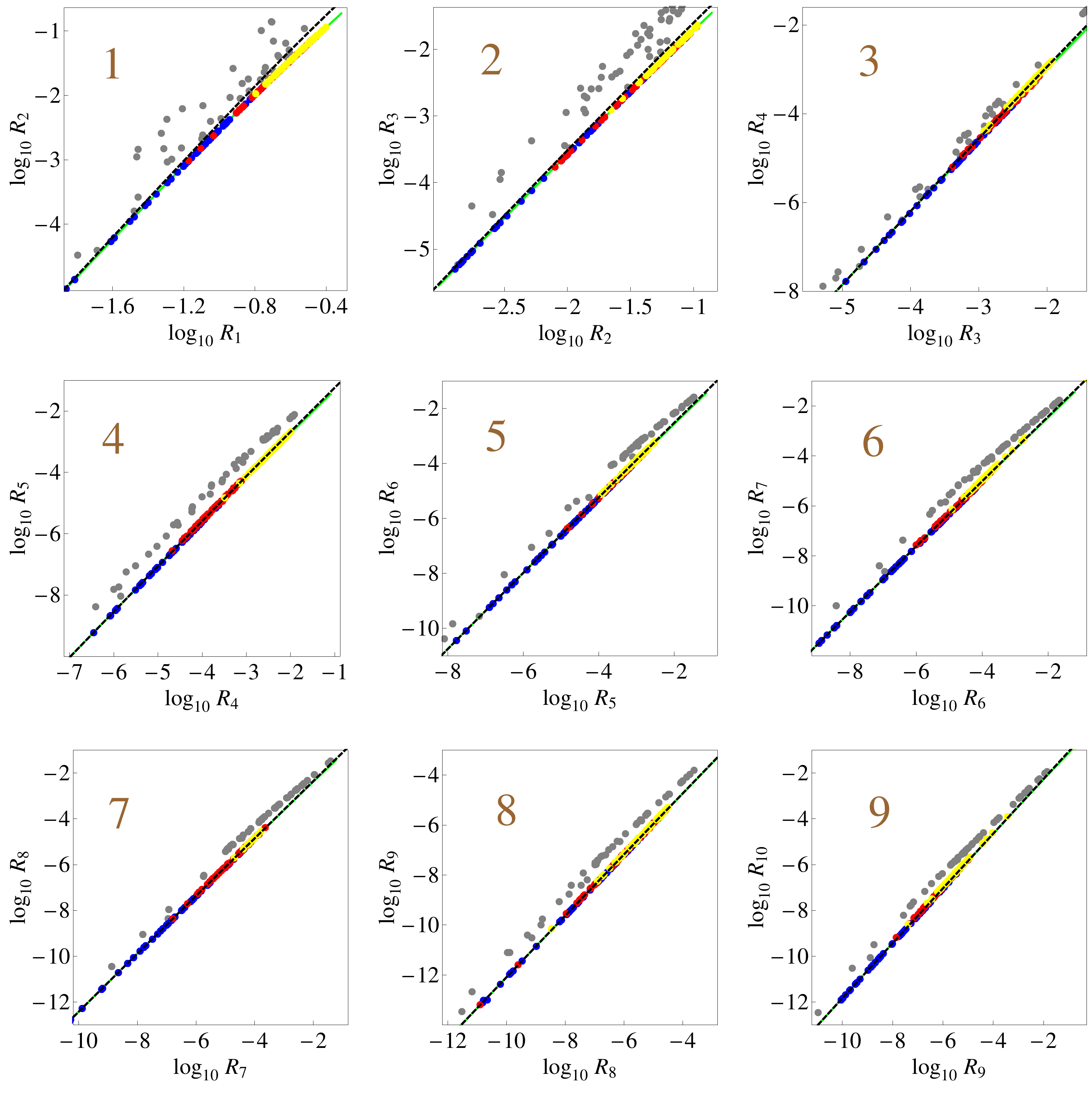} 
\caption{The Figure depicts thresholds of quantum non-Gaussianity by solid green lines.  The dashed black lines appropriate to the approximative parametrizing (\ref{RnEq}). For higher order, the approximations are practically the same as their exact counterparts. The thresholds can not be overcome by any Gaussian state of light, which was verified by Monte - Carlo method. The closest 50 random attempts are depicted for single (blue), two (red) and three (yellow) modes of Gaussian light. The total number of runs were $10^5$ (single mode), $10^6$ (two modes) and $10^7$ (three modes). The gray points show a random arrangement of points generated by $50$ runs for a single Gaussian mode.}
\label{fig.depthSupplement}
\end{figure*}

The thresholds of quantum non-Gaussianity were derived under the assumption that squeezing of an optimal Gaussian state is minimal in the direction of its amplitude, i.\ e.\ $\phi=0$. Also, we presume that optimizing over a single mode of a Gaussian light suffices to obtain thresholds covering even multi-mode Gaussian light. Those assumptions enable to gain a numerical solution of this task. However, there is no general analytic or semi-analytic proof of their correctness and therefore the only way of their verification is the Monte Carlo method. A general $M$-mode Gaussian light is fully described by $3 M$ parameters. The range of parameters in $i$-th mode was set to $\beta_i^2 \in ( 0; 2 n )$, $V_i \in \left(\frac{1}{n+2}; 1 \right)$ and $\phi_i \in (0; 2 \pi )$, where $i$ goes from one to $M$ and $n$ corresponds to the order of the criterion that is being verified. The number of runs the method has to perform to become reliable increases with the number of simulated modes. From this limitation, we chose the maximal number of modes in our test to be three. The thresholds were checked by $10^5$, $10^6$ and $10^7$ runs for $1$, $2$ and $3$ modes, respectively. The result is demonstrated in Fig.~\ref{fig.depthSupplement}, which depicts 50 points closest to the thresholds. All these tests indicate that $M$-mode Gaussian light can reach the threshold only if a single mode is occupied by photons and the rest $n-1$ modes are vacuum.

\begin{figure*}
\includegraphics[width=.7\linewidth]{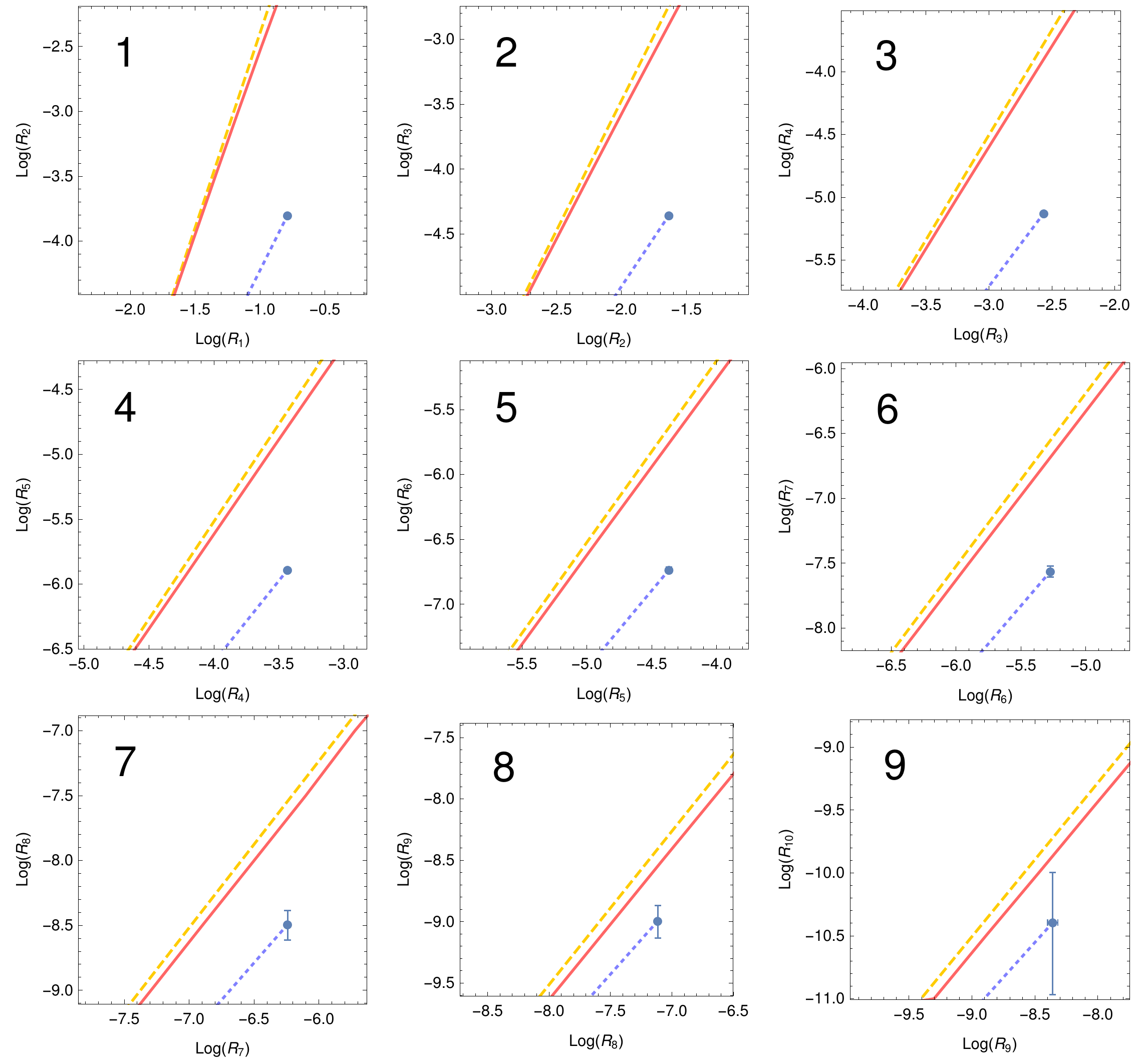}
\caption{QNG thresholds with experimental data for heralded 1-9 photons. The probabilities $R_n$ and $R_{n+1}$ for generated $n$-photon states were measured on a balanced $(n+1)$-channel detector (blue points for $n=1-9$). Solid red lines represent the respective QNG criteria, while dashed orange lines are an approximation (Eq.\ \ref{RApp}). Dotted blue lines represent the path of the points if the states become further attenuated. Both vertical and horizontal errorbars are shown for all data points; in some cases they are smaller than point size. For more details, see the main text.}
\label{fig.dataSupplement}
\end{figure*}

The limit of very attenuated states $R_{n+1}\ll 1$ is experimentally relevant and therefore interesting to discuss. For this class of states, the criteria can be simplified approximately by analytical formulas. Let us denote the probability of $n$ photons in a state $\rho$ by $P_n=\langle n \vert \rho \vert n \rangle$. Interestingly, a source of Gaussian light can generate a state with $P_{n+2} \gg P_{n+1} \sim P_{n+3}$. The detection probabilities of this state are mainly contributed by
\begin{eqnarray}
R_n &\approx & T_{n,n} P_n \nonumber \\
R_{n+1} &\approx & T_{n+1,n+1} P_{n+1} \nonumber \\
 &+&T_{n+1,n+2} P_{n+2}+T_{n+1,n+3} P_{n+3},
\label{RnProp}
\end{eqnarray}
where $T_{n,m}$ is an element of a matrix denoting the probability of $n$ simultaneous detections caused by $m$ photons. In this approximation, the success events are caused mainly by $n$ photons, whereas error events are rendered significantly even by $n+3$ photons, due to the assumption that probability of $n+1$ photons in the Gaussian state is surpassed. The matrix $T$  has elements
\begin{equation}
T_{n,m}=1+\sum_{k=1}^{n}{n \choose k}(-1)^k\left(1-\frac{k}{N}\right)^m,
\label{Tnm}
\end{equation}
where $N$ is the total number of channels in the multiplex detector. This expression can be written in a more convenient form, when the summation is added up:
\begin{eqnarray}
T_{n,m}&=&0\mbox{ ... }n>m \nonumber \\
T_{n,n}&=&\frac{n!}{N^n} \nonumber \\
T_{n,n+1}&=&\frac{(n+1)!}{N^n}\left(1-\frac{n}{2 N}\right)\nonumber \\
T_{n,n+2}&=&\frac{1}{24}\frac{(n+2)!}{N^{n+2}}(n+3 n^2-12 n N+12 N^2).
\end{eqnarray}
The derivation of these relations is lengthy, but their correctness can be verified by (\ref{Tnm}).

The photon statistic of a squeezed coherent state gains \cite{Yuen,Stoler}
\begin{eqnarray}
P_n&=&\frac{2\sqrt{V}}{(1+V)n!} H_n^2\left(\sqrt{ V/(1-V^2)/2}\beta\right)  \nonumber \\
& \times & \left[\frac{1-V}{2(1+V)}\right]^n\exp \left[-\frac{\beta^2 V}{2(1+V)} \right],
\label{GStatistic}
\end{eqnarray}
where $H_n$ is a Hermite polynomial of order $n$.
This statistic appropriates to states with $\phi=0$. It means the quadrature variance is minimal in the direction of the amplitude of the coherent state. 
The assumed condition $P_{n+1}\ll P_{n+2}$ is justified if
\begin{equation}
 H_{n+1}\left(\frac{\beta \sqrt{V}}{2\sqrt{1-V}}\right)=0.
 \end{equation}
Let us identify the ratio in the argument with $x=\frac{\beta \sqrt{V}}{2\sqrt{1-V}}$ and connect
\begin{eqnarray}
\beta^2 V&=&4 x^2 t \nonumber \\
V&=&1-t,
\label{para}
\end{eqnarray}
where $t$ is a parameter.
The equation $H_{n+1}(x)=0$ gives raise to a relation between squeezing $V$ and amplitude $\beta$. Taking into account the assumption (\ref{RnProp}), parametrization (\ref{para}), photon statistic of Gaussian states (\ref{GStatistic}) expanded according to $t$ and setting the total number of channels $N=n+1$, the probabilities of clicks yield approximations
\begin{eqnarray}
R_n &\approx&\frac{1}{2}\frac{H_n^2(x)}{\left[4(n+1)\right]^n}t^n(2+n t) \nonumber \\
R_{n+1}& \approx &\left[12(1+n)+6 (1+n)(2+n)t-x^2 (2+3 n)t \right] \times \nonumber \\ &\times &\frac{H_{n}^2(x)}{3\times 2^5\left[4(n+1)\right]^{n}}t^{n+2},
\label{RnEq}
\end{eqnarray}
where $x$ is such that $H_{n+1}(x)=0$ and $t > 0$ is a parameter. If there are more than one $x$ satisfying the condition, only the $x$ giving the greatest value of $H_n(x)$ is considered.

Fig.~\ref{fig.depthSupplement} compares this approximation with the exact solution. For small $t \ll 1$, some members can be omitted and the exclusion of $t$ leads to an analytical approximation for threshold success probability
\begin{equation}
R_{n}^{n+2}=H_n^{4}(x) \left[ \frac{R_{n+1}}{2 (n+1)^3}\right]^{n}.
\label{RApp}
\end{equation}

This relation is suitable for $\sqrt[n+2]{R_{n+1}}\ll 1$. Very interestingly, the approximative parametrizing of the threshold (\ref{RnEq}) holds precisely even for $t \lesssim 1$, especially for large $n$.

\subsection{Experimental parameters}

The 2-ns coincidence window was selected as a good trade-off between two negative effects. On one hand, the combined jitter of all the detectors causes effective detection loss for narrow coincidence windows. That decreases QNG depth. On the other hand, as the coincidence window broadens, undesirable higher photon-number contributions add excess noise to the multi-photon state, which also decreases QNG depth. Although the lower limit of the coincidence window depends on the number of single-photon detectors and therefore on the QNG criterion order, we decided to use a fixed value for all measurements. This way, we kept all the heralded single-photon states identical in terms of their photodistribution, independently on the measurement they were subjected to.

For detection, we used two time-to-digital converters, each having 8 channels with the resolution of 81 ps/time bin. Since we needed 11 channels in total, we used two modules and synchronized them with a shared periodic signal at 100 kHz. This frequency was chosen to compensate the measured relative clock drift $10^5$ time bins/second.

\subparagraph{Data availability}
The data that support the findings of this study are available from the corresponding author upon request. 

\subparagraph{Conflict of interest:}

The authors declare no conflict of interest.

\section{Author Contributions}
L.~L. and R.~F. derived the QNG criteria and provided the theoretical analysis. I.~S. built the SPDC source and performed the measurement and data analysis. J.~H. built the multichannel detector. M.~J., M.~Mikov\' a and M.~Mi\v cuda participated in the experimental work. M.~J. closely supervised and coordinated the experiment. I.~S., R.~F. and L.~L. wrote the manuscript and all authors were involved in creating and revising the manuscript. R.~F. initiated and coordinated the project.

\section{Acknowledgements}
L.~L., J.~H.,  M.~J. and R.~F. acknowledge the financial support of the Czech Science Foundation (GB14-36681G). I.~S. and M.~Mikov\' a acknowledge the financial support of Palack\' y University (IGA-PrF-2016-009, IGA-PrF-2017-008).

\bibliography{references_arXiv_2}

\end{document}